# Achieving accurate entropy and melting point by *ab initio* molecular dynamics and zentropy theory: Application to fluoride and chloride molten salts

Shun-Li Shang,[1,*] Nigel L. E. Hew,[1] Rushi Gong,[1] Cillian Cockrell,[2] Paul A. Bingham,[3] Xiaofeng Guo,[4] Jingjing Li,[5] Qi-Jun Hong,[6] and Zi-Kui Liu[1]

1. Department of Materials Science and Engineering, The Pennsylvania State University, University Park, PA 16802, USA
2. Nuclear Futures Institute, Bangor University, Bangor LL57 1UT, UK
3. School of Engineering and Built Environment, Sheffield Hallam University, Sheffield S1 1WB, UK
4. Department of Chemistry, Washington State University, Pullman, Washington, 99164, USA
5. Department of Industrial and Manufacturing Engineering, The Pennsylvania State University, University Park, PA 16802, USA
6. Materials Science and Engineering, School for Engineering of Matter, Transport, and Energy, Arizona State University, Tempe, Arizona 85285, USA

*E-mail: sus26@psu.edu (S. L. Shang)

## Abstract

We have recently developed a breakthrough methodology for rapidly computing entropy in both solids and liquids by integrating a multiscale entropy approach (known as zentropy theory) with molecular dynamics (MD) simulations. This approach enables entropy estimation from a single MD trajectory by analyzing the probabilities of local structural configurations and atomic distributions, effectively addressing the long-standing challenge of capturing configurational entropy, particularly for liquid. Here, we demonstrate the power of this method by predicting entropies, enthalpies, and melting points of 25 binary and ternary fluoride- and chloride-based molten salts using ab initio MD (AIMD) simulations. The remarkable agreement between our predictions and experimental data underscores the potential of this approach to transform computational thermodynamics, offering accurate, efficient, and direct predictions of thermodynamic properties across both solid and liquid phases.

## Highlights

- Rapid and accurate computation of entropy in both solids and liquids
- Zentropy-based predictions of configurational entropy using a single MD trajectory
- Accurate prediction of entropy, enthalpy, and melting point
- Exploration of 25 fluoride and chloride molten salts

## 1 Introduction

Unlike solids and gases, thermodynamic properties of liquids, particularly entropy, remains poorly understood due primarily to the complex interplay of strong interatomic interactions and dynamic disorder, which prevents the establishment of long-range order [1,2]. This makes it difficult to define a reference state necessary for entropy calculations. Consequently, understanding and modeling liquids rely heavily on classical molecular dynamics (MD) or ab initio molecular dynamics (AIMD) simulations [3]. MD simulations have proven effective in predicting enthalpy ($H$) [4], however, it is widely accepted that entropy ($S$) cannot be directly or accurately determined for liquids [5]. As a result, thermodynamic properties that depend on liquid entropy, such as Gibbs energy, liquidus and solidus temperatures, and the present focus of melting point ($T_\mathrm{m}$), remain challenging to compute with precision.

Given the prevailing belief that entropy cannot be accurately determined from MD simulations, most state-of-the-art methods for predicting melting points circumvent direct entropy calculations [6,7]. Instead, alternative strategies have been developed within the MD framework. For example, free energy methods, which employ sophisticated approaches to calculate liquid free energy like thermodynamic integration, the two-phase thermodynamic method, and the Widom particle insertion method [7]; the fast-heating method, where the solid is rapidly heated until it melts [8]; the voids method, which involves removing atoms to induce melting [9]; and solid-liquid coexistence methods, such as the interface pinning method [10,11] and especially the SLUSCHI (Solid and Liquid in Ultra Small Coexistence with Hovering Interfaces) method developed by Hong and van de Walle [7]. SLUSCHI determines melting points by analyzing enthalpy changes as a function of temperature in small supercells containing coexisting solid and liquid phases, offering a computationally efficient alternative to traditional large-scale coexistence simulations [7]. Beyond theoretical and simulation-based methods, machine learning (ML) techniques have also been developed to predict melting points. For example, Hong et al. [12] built a graph neural network model to predict melting points across a diverse range of materials from ancient minerals to novel compounds. Varnet et al. [13] applied various ML algorithms such as neural networks (NN), support vector machines (SVM), and $k$-nearest neighbors ($k$NN) to predict melting points of organic materials.

Recently, Hong and Liu [14] introduced a groundbreaking methodology for rapid and accurate entropy calculations for both solids and liquids by integrating a multiscale entropy approach known as zentropy theory [15–17] with MD or AIMD simulations; see details in Sec. 2.1. This zentropy-based approach addresses the long-standing challenge of entropy calculations across a wide range of materials — including perfect crystals, amorphous phases, and liquids — by decomposing total entropy into configurational, vibrational, and thermal electronic components [14]. This methodology has been successfully applied to predict melting points for both pure elements (e.g., Al) and complex compounds (e.g., $ZrO_2$, HfC, $HfC_{0.88}$, and $HfC_{0.5}N_{0.38}$) [14], by directly equating the Gibbs energies of the solid and liquid phases, i.e.,

$$T_{\mathrm{m}} = \Delta H_{\mathrm{m}} / \Delta S_{\mathrm{m}} \quad \textbf{Eq. 1}$$

where $\Delta H_{\mathrm{m}}$ and $\Delta S_{\mathrm{m}}$ are the melting (fusion) enthalpy and entropy, respectively, and $T_{\mathrm{m}}$ represents the melting point. For solution phases or non-stoichiometric compounds, $T_{\mathrm{m}}$ corresponds to the $T_0$ temperature, where two phases with the same composition have equal Gibbs energy. Compared to conventional methods such as SLUSCHI [7], the zentropy-based approach offers significant computational advantages besides accurate calculations of entropy. It requires smaller supercells (typically half the size used in SLUSCHI) and fewer MD temperature points (one is enough for most cases; see Sec. 3), making its orders of magnitude faster while maintaining high accuracy, compared to traditional coexistence-based simulations like SLUSCHI.

In the present work, the zentropy-based methodology [14] is employed to predict entropies, enthalpies, and melting points for twenty-five binary and ternary fluoride and chloride molten salts using AIMD simulations. The salts investigated include cations from five alkali metals (Li, Na, K, Rb, and Cs), five alkali earth metals (Be, Mg, Ca, Sr, and Ba), and two transition metals (Zn and Y). A complete list of the studied salts is provided in Table 1, the supplementary Table S 1 and the supplementary Excel file. The present predictions demonstrate remarkable accuracy compared with experiments, highlighting the unique capability of zentropy theory to accurately compute entropy and entropy-related properties in complex molten salt systems.

## 2 Methodology

Table 1 provides an overview of the computational methods and key equations used to predict the melting point, enthalpy, entropy, and heat capacity for both solid and liquid phases. Detailed methodologies are presented in the following subsections.

### 2.1 Zentropy theory for solids and liquids

We have developed a multiscale entropy framework, recently termed zentropy theory [16], with "$z$" standing for the partition function from the German word *zustandssumme*, meaning "sum over states" [15–17]. Zentropy theory enables coarse graining of entropy using nested formulae, spanning various scales from the observed system down to the quantum scale [15–18]; see Figure 1. In this framework, the total entropy ($S_{\text{tot}}$) of a system is computed by combining two perspectives: a bottom-up summation of quantum entropy $S^k$ of each configuration $k$, and a top-down statistical entropy among configurations ($S_{\text{conf}}$),

$$S_{\text{tot}} = \sum_{k=1}^{m} p^k S^k - k_B \sum_{k=1}^{m} p^k \ln p^k, \qquad \textbf{Eq. 2}$$

where $m$ is the number of configurations, $k_B$ the Boltzmann constant, and $p^k$ the probability of configuration $k$. It is important to note that **Eq. 2** accounts for all entropy contributions, including kinetic components such as translational and rotational entropies. Taking reversible Brownian motion as an illustrative example [15], when the probability of an atom jumping to a neighboring vacant site becomes statistically significant, an additional entropy term arises from the 2nd summation in **Eq. 2**. This internal reversible process is driven by thermal energy absorbed from the surroundings, thereby increasing the system's entropy. The transition between various states is included in the Gibbs statistical mechanics where Gibbs "imagined a great number of systems of the same nature, but differing in the configurations and velocities which they have at a given instant, and differing not merely infinitesimally, but it may be so as to embrace every conceivable combination of configuration and velocities" [19].

Usually each $S^k$ in **Eq. 2** includes both vibrational (vib) and thermal electronic (ele) contributions, i.e., $S^k = S_{\text{vib}}^k + S_{\text{ele}}^k$, according to the density functional theory (DFT) based quasiharmonic approximation (QHA) [20]. Substituting this into **Eq. 2** yields,

$$S_{\text{tot}} = S_{\text{vib}} + S_{\text{ele}} + S_{\text{conf}}, \qquad \textbf{Eq. 3}$$

where $S_{\text{vib}} = \sum_{k=1}^{m} p^k S_{\text{vib}}^k$, $S_{\text{ele}} = \sum_{k=1}^{m} p^k S_{\text{ele}}^k$, and $S_{\text{conf}} = -k_B \sum_{k=1}^{m} p^k \, ln p^k$. It is important to note that in classical statistical mechanics, $S^k = 0$ is assumed, which is valid only for a pure quantum state. However, at finite temperatures, DFT predicts $S^k > 0$ [20], justifying the use of **Eq. 2** and **Eq. 3**. In zentropy theory, the canonical partition function of the system is reformulated by replacing the internal energy of each configuration in Gibbs's classical statistical mechanics with its Helmholtz energy [15–18],

$$Z = e^{-\frac{F}{k_B T}} = \sum_{k=1}^{m} Z^k = \sum_{k=1}^{m} e^{-\frac{F^k}{k_B T}} = \sum_{k=1}^{m} e^{-\frac{E^k - TS^k}{k_B T}}, \quad \textbf{Eq. 4}$$

where $Z$ and $Z^k$ are the partition functions with $p^k = Z^k/Z$, $F$ and $F^k$ are the Helmholtz energies of the system and the configuration $k$, respectively, $E^k$ is the internal energy, and $T$ the absolute temperature. For stable and metastable configurations, $F^k$ as well as $E^k$ and $S^k$ can be predicted by the DFT-based QHA [20]. The total Helmholtz energy is then given by,

$$F_{\text{tot}} = -k_B T ln Z = \sum_{k=1}^{m} p^k F^k - TS_{\text{conf}}. \quad \textbf{Eq. 5}$$

It is important to note that all emergent behaviors of a system arise from the statistical competition among its configurations, governed by their individual intrinsic properties, i.e., their $E^k$ and $S^k$ and their responses to external stimuli. Zentropy theory has successfully predicted emergent properties in solid phases using $p^k$, $S_{\text{conf}}$, $F_{\text{tot}}$, and their 1st and 2nd derivatives, for example, critical point, heat capacity, positive and negative thermal expansion, and phase diagram in bcc Fe [21], fcc Ce [22,23], fcc Ni [24], $Fe_3Pt$ [25], $YNiO_3$ [26], $SmNiO_3$ [27], $BaFe_2As_2$ [28], $Cu_2ZnSnS_4$ [29], and FeNiCoCr medium entropy alloy [30]. Notably, these results were achieved using DFT-based calculations with small supercells and a limited number of independent symmetry-broken configurations, typically fewer than a few hundred, demonstrating the efficiency and scalability of zentropy theory.

Unlike solids, which often exhibit well-defined configurations, distinguishing configurations in liquids is inherently challenging. To address this, Hong and Liu [14] proposed a method to identify configurations in both solids and liquids by analyzing the probability of local structural arrangements and atomic distributions. Following a single MD trajectory, configurational entropy for both solids and liquids can be computed by analyzing the nearest-neighbor environments of atomic pairs [14],

$$S_{\text{conf}} = -\left(1-\frac{\delta}{2}\right) k_B \sum_{n=0}^{n_{\max}} p^n \ln p^n. \qquad \textbf{Eq. 6}$$

Here, $\delta = 1$ if the atomic pair consists of identical elements, and $\delta = 0$ otherwise. $p^n$ represents the probability of an atom having $n$ nearest neighbors, up to a maximum of $n_{\max}$.

For multicomponent systems such as $\text{A}_{x_\text{A}}\text{B}_{x_\text{B}}\text{C}_{x_\text{C}}$ with $x_\text{A} + x_\text{B} + x_\text{C} = 1$, the total configurational entropy per atom is computed as a weighted sum of the individual elemental contributions [14],

$$S_{\text{conf}} = x_\text{A} S_{\text{conf}}(\text{A}) + x_\text{B} S_{\text{conf}}(\text{B}) + x_\text{C} S_{\text{conf}}(\text{C}), \qquad \textbf{Eq. 7}$$

where $S_{\text{conf}}(\text{A})$, $S_{\text{conf}}(\text{B})$, and $S_{\text{conf}}(\text{C})$ are the configurational entropies associated with elements A, B, and C, respectively. Various options are available for choosing the nearest-neighbor pairs. For example, $S_{\text{conf}}(\text{A})$ can be computed based on $p^n$ of the A-A, A-B, or A-C pairs. Herein, the following guidelines are applied based on our prior work [14] and current practices,

(1) Primary (1st): Pairs between the same elements and within the same sublattice (e.g., A-A for liquid phase or within the A sublattice for solid phase).
(2) Secondary (2nd): Pairs between heavier elements.
(3) Tertiary (3rd): Pairs between a light element and a previously selected heavy element.

In the present work, the first guideline is used for liquids when the corresponding solids, simulated at the same temperature, exhibit negligible $S_{\text{conf}}$ (e.g., < 0.1 J/K·mol-atom). The second and the third guidelines are also considered for liquids when the corresponding solids have a non-negligible $S_{\text{conf}}$ (e.g., > 0.1 J/K·mol-atom), and the average $S_{\text{conf}}$ is adopted for liquids in these cases, e.g., $S_{\text{conf}} = \left[S_{\text{conf}}^{(1^{\text{st}})} + S_{\text{conf}}^{(3^{\text{rd}})}\right]/2$ for binary compounds. Note that the first guideline is consistently applied to solids.

In addition to $S_{\text{conf}}$, the vibrational entropy $S_{\text{vib}}$ (see **Eq. 3**) is obtained from the same MD simulations via phonon analysis using velocity auto-correlation functions, as detailed in [14,31]. The thermal electronic entropy $S_{\text{ele}}$ (see **Eq. 3**) is computed by averaging electronic entropy values along the DFT-based MD trajectory [14]. However, $S_{\text{ele}}$ is only non-negligible for conductors and this term equals zero for the present work since both fluoride and chloride salts are insulators.

The zentropy-based methodology, based on MD trajectory, has been implemented in the SLUSCHI package [7], accessible via the mds (MD entropy) or mds_lmp (integration with the LAMMPS code) commands [32]. The updated SLUSCHI code was employed in the present work.

## 2.2 Details of AIMD simulations

All DFT-based AIMD simulations in the present work were performed using the Vienna Ab initio Simulation Package (VASP) [33]. The ion-electron interactions were described using the projector augmented wave (PAW) method [34], while the exchange-correlation (X-C) functionals were described using both the local density approximation (LDA) [35,36] and the generalized gradient approximation (GGA) by Perdew, Burke, and Ernzerhof (PBE) [37]. Electron configurations for each element were consistent with those used by the Materials Project [38]. In addition, test calculations were also performed using the X-C functional of metaGGA, i.e., the $r^2$SCAN [39] for two salts of $BaF_2$ and $SrF_2$.

AIMD simulations via VASP were managed by SLUSCHI [7,32], employing both the isobaric-isothermal (*NPT*) and canonical (*NVT*) ensembles, where *N*, *P*, *T*, and *V* represent the number of atoms, pressure, temperature, and volume, respectively. A special *k*-point mesh of (¼ ¼ ¼) was used for more accurate results in all calculations, than the Γ-centered mesh. The plane-wave cutoff energy for each compound was set to the maximum of the default values (using "PREC = Normal" in VASP), and further increased to high precision ("PREC = High") during Pulay stress corrections [40]. Electronic energy convergence was set as $10^{-4}$ eV/supercell. Temperature control was implemented using the Nosé-Hoover chain formalism, while pressure control was achieved by adjusting the volume every 80 ionic steps during *NVT* simulations, according to the average pressure [41]. All simulations were conducted at zero external pressure, meaning the Helmholtz energy is equivalent to the Gibbs energy in the present work. For each temperature, AIMD simulations were performed for approximately 280 runs for liquids and 250 runs for solids, with each run consisting of 80 ionic steps and a time step of 3 fs. Notably, melting point predictions stabilized after approximately 100 AIMD runs, with temperature fluctuations around 1%; see supplementary Figure S 1 for tests for KCl using LDA at 1100 K. Figure S 1a illustrates the convergence of the liquid phase temperature for KCl using LDA. It shows that after 10 AIMD runs, the simulations reach the target temperature of 1100 K, with a prediction of 1100 ± 10 K. Figure S 1b confirms that the predicted melting point exhibits clear convergence after 100 AIMD runs. All other AIMD settings in the present work followed SLUSCHI's default recommendations [7,32].

For each of the 25 fluoride and chloride salts studied, supplementary Table S 1 and the accompanying Excel file provide detailed information, including crystallographic structure, space group, supercell size (ranging from 88 to 128 atoms), cutoff distances used to generate the supercells, and AIMD simulating temperatures. The variation in both melting enthalpy ($\Delta H_{\mathrm{m}}$) and melting entropy ($\Delta S_{\mathrm{m}}$) with respect to temperature near the melting point ($T_{\mathrm{m}}$) is relatively small. As a result, AIMD simulations conducted at one or two temperatures around $T_{\mathrm{m}}$ are generally sufficient to estimate $\Delta H_{\mathrm{m}}$ and $\Delta S_{\mathrm{m}}$, and subsequently determine $T_{\mathrm{m}}$ for most cases. The selected AIMD temperatures can be guided by experimentally measured melting points or machine learning predictions [12]. Initial liquid configurations were generated by heating the system two or three times the target simulation temperature during the first five AIMD runs. After AIMD simulations, thermodynamic properties were analyzed by the mds command in SLUSCHI, with the first 50 AIMD runs discarded to ensure equilibration.

## 3 Results and discussion

Table 2 summarizes the predicted melting point ($T_{\mathrm{m}}$), melting enthalpy ($\Delta H_{\mathrm{m}}$), and melting entropy ($\Delta S_{\mathrm{m}}$) for 25 fluoride and chloride molten salts by analyzing AIMD trajectories; see Sec 2.1. For salts simulated at multiple temperatures, the reported values of $T_{\mathrm{m}}$, $\Delta H_{\mathrm{m}}$, and $\Delta S_{\mathrm{m}}$ are derived from linear fits to the simulated results. Additional equilibrium properties for solids and liquids at zero external pressure are provided in the supplementary Excel file, including volume, enthalpy (without kinetic energy of $1.5k_{\mathrm{B}}T$), vibrational entropy, configurational entropy, and melting point, along with experimental data from Janz [42] and the SGTE substance database (SSUB5, hereafter referred to as SSUB) [43]. It is worth noting that Janz [42] and SSUB [43] report different melting points for five salts (RbCl, RbF, $SrCl_2$, $ZnCl_2$, and especially $YF_3$, see Table 2). In this section, we present first a detailed case study of KCl (Sec.3.1), followed by a comprehensive analysis of 25 molten salts (Sec. 3.2 and Sec. 3.3). It is worth noting that heat capacity predictions are demonstrated only for KCl. For other salts, heat capacity is not evaluated due to the limited number of AIMD simulation temperatures, as the present focus is on melting point determination.

### 3.1 Case study of KCl

To compute the configurational entropy, $S_{\mathrm{conf}}$, of liquid KCl using **Eq. 6**, Figure 2 presents the predicted probability distributions $p^n$ as a function of the number of nearest neighbors *n* for both K and Cl after AIMD simulations using LDA and 88-atom supercell at 1100 K. Here, the sum of all $p^n$ equals unity for each atomic pair. Figure 2 reveals the following distributions in liquid KCl. The K–K pairs exhibit nearest-neighbor counts ranging from *n* = 11 to 20, with the highest probability at $p^{n=15}$ (Figure 2a). The K-Cl pairs (with K as the central atom) range from *n* = 3 to 7, peaking at $p^{n=5}$ (Figure 2b); the Cl-K pairs (with Cl as the central atom) range from *n* = 3 to 8, with the maximum probability being $p^{n=6}$ (Figure 2c), and the Cl-Cl pairs range from *n* = 9 to 17, with the highest probability at $p^{n=13}$ (Figure 2d). For each element (K or Cl), two approaches are used to compute $S_{\mathrm{conf}}$ in liquids. For instance, $S_{\mathrm{conf}}$(K) = 6.94 J/K·mol-atom based on the K-K pairs (with *n* ranging from 12 to 20), and $S_{\mathrm{conf}}$(K) = 5.27 J/K·mol-atom based on the K-Cl pairs (with *n* ranging from 3 to 7). Since $S_{\mathrm{conf}}$ = 0 for solid KCl at 1100 K (not shown), we select $S_{\mathrm{conf}}$(K) = 6.94 J/K·mol-atom following the first guideline (cf., Sec. 2.1). Similarly, $S_{\mathrm{conf}}(\mathrm{Cl})$ = 7.14 J/K·mol-atom is selected based on the Cl-Cl pairs; see Figure 2. The average $S_{\mathrm{conf}}$ of liquid KCl at 1100 K is therefore 7.04 J/K·mol-atom by averaging $S_{\mathrm{conf}}$(K) and $S_{\mathrm{conf}}(\mathrm{Cl})$ based on the composition of KCl. Regarding vibrational entropy, the predicted $S_{\mathrm{vib}}$ = 79.5 J/K·mol-atom for liquid KCl at 1100 K. The total entropy is hence 86.6 J/K·mol-atom, which is in good agreement with the experimental entropy of 91.4 J/K·mol-atom from SSUB [43]. Figure S 2 shows a further entropy comparison of KCl between the present predictions by both LDA and GGA and those from the SSUB. It indicates that GGA predicts more accurate total entropies for both solid and liquid KCl. While the predicted entropy differences ($\Delta S$) between solid and liquid KCl are similar (e.g., 10.7 J/K·mol-atom by LDA and 10.3 J/K·mol-atom by GGA at 1100 K) and are compatible with the constant $\Delta S$ value of 12.6 J/K·mol-atom from SSUB [43].

Figure 3 presents the predicted vibrational ($S_{\mathrm{vib}}$) and configurational ($S_{\mathrm{conf}}$) entropies of solid and liquid KCl as a function of temperature, based on AIMD simulations using LDA and an 88-atom supercell (see supplementary Excel file for details). It shows the near-zero $S_{\mathrm{conf}}$ values for solids, while $S_{\mathrm{conf}}$ reaches approximately 7.0 (the unit is J/K·mol-atom for entropy) for liquids near the melting point (measured $T_{\mathrm{m}}$ = 1044 K [42,43], see also Figure S 2). This yields a configurational entropy difference $\Delta S_{\mathrm{conf}}$ = 7.0 between the liquid and solid phases. Although the absolute $S_{\mathrm{vib}}$ values are high, exceeding 70 J/K·mol-atom for both phases, their difference near the melting point is relatively small, with $\Delta S_{\mathrm{vib}}$ = 3.6. Using the predicted $\Delta H_{\mathrm{m}}$ = 11.22 kJ/mol-atom at 1100 K and **Eq. 1**, the predicted $T_{\mathrm{m}}$ is 1143 K (when $\Delta S_{\mathrm{m}} = \Delta S_{\mathrm{vib}} + \Delta S_{\mathrm{conf}}$) or 3347 K (when $\Delta S_{\mathrm{m}} = \Delta S_{\mathrm{vib}}$). These results highlight the critical role of $\Delta S_{\mathrm{conf}}$ in predicting melting point.

It is also noteworthy that the predicted $T_{\mathrm{m}}$ depends on several factors, including (1) the X-C functional, for example, GGA predicts $T_{\mathrm{m}}$ = 927 K based on AIMD simulations at 1000 K and 1100 K for KCl; (2) simulation temperature, for example, LDA yields $T_{\mathrm{m}}$ = 1114 K, 1143 K, and 1127 K based on simulations at 1000 K, 1100 K and 1200 K for KCl, respectively; and (3) supercell size, for example, LDA predicts $T_{\mathrm{m}}$ = 1053 K using a 92-atom supercell for KCl. Considering both LDA and GGA predictions using the 88-atom supercell, the present work shows the predicted $T_{\mathrm{m}}$ = 1028 ± 102 K (i.e., 1130 from LDA and 927 from GGA, cf., Table 2 and supplementary Excel file), agreeing well with the measured 1044 K of KCl [42,43]. The higher $T_{\mathrm{m}}$ predicted by LDA and the lower $T_{\mathrm{m}}$ from GGA reflect a well-known DFT trend, i.e., LDA typically overestimates bonding strength and GGA underestimates it [44], while $T_{\mathrm{m}}$ is proportional to bonding strength [45].

As another example, the specific heat capacity at constant pressure, $C_{\mathrm{P}}$, can be predicted from the temperature dependence of enthalpy or entropy as follows,

$$C_{\mathrm{P}} = \left(\frac{\partial H_{\mathrm{tot}}}{\partial T}\right)_{\mathrm{P}}, \qquad \textbf{Eq. 8}$$

$$C_{\mathrm{P}} = T\left(\frac{\partial S_{\mathrm{tot}}}{\partial T}\right)_{\mathrm{P}} = T\left(\frac{\partial S_{\mathrm{vib}}}{\partial T} + \frac{\partial S_{\mathrm{conf}}}{\partial T}\right)_{\mathrm{P}}, \qquad \textbf{Eq. 9}$$

where $H_{\mathrm{tot}}$ is the total enthalpy, including both potential energy (from AIMD) and kinetic energy (= $1.5k_{\mathrm{B}}T$) at a given temperature. Based on AIMD simulations using LDA at 1000 K, 1100 K, and 1200 K, the predicted $C_{\mathrm{P}}$ values for solid and liquid KCl are plotted in Figure 4, compared with the $C_{\mathrm{P}}$ values from SSUB [43]. The results indicate that the liquid $C_{\mathrm{P}}$ is higher than the solid $C_{\mathrm{P}}$, consistent across both enthalpy- and entropy-based predictions. Including $S_{\mathrm{conf}}$ (via **Eq. 9**), the predicted $C_{\mathrm{P}}$ of liquid (represented by red closed circles in Figure 4) increases by about 10% compared to the $S_{\mathrm{vib}}$-only predictions, aligning closely with the experimental value from SSUB.

Moreover, the inclusion of configurational entropy explains the observed jump in $C_\mathrm{P}$ at the melting point, making the transition from solid to liquid; see Figure 4.

The present KCl case study demonstrates that zentropy-based MD simulations enable accurate predictions of entropy and entropy-related properties, such as melting points and heat capacity.

### 3.2 Melting points of 25 fluoride and chloride salts

Figure 5 summarizes the predicted melting points ($T_\mathrm{m}$) for 25 fluoride and chloride salts using **Eq. 1** in terms of AIMD simulations with both LDA and GGA, compared to experimental data from the SSUB database [43]. The digital data are provided in Table 2 and the supplementary Excel file. On average, the predicted $T_\mathrm{m}$ values (represented by red open circles in Figure 5 via averaging LDA and GGA results) align reasonably well with experimental data, yielding a mean absolute error of 92 K or 8 %. As expected, LDA generally predicts higher $T_\mathrm{m}$ values than GGA, except for CsF, $BaCl_2$, $BaF_2$, $BeCl_2$, $CaCl_2$, $SrF_2$, $YCl_3$, and $Li_2BeF_4$. Notably, better agreement is mainly achieved for salts with negligible $S_\mathrm{conf}$ values in their solids near the melting points (see supplementary Excel file), for example, all the NaCl-type alkali metal halides which have an average error of 46 K or 5%. In contrast, larger deviations are observed for salts with significant $S_\mathrm{conf}$ in their solids. The top three largest errors are found for $BeF_2$ (error of 202 K or 24%), $SrF_2$ (330 K or 19%), and $BaF_2$ (277 K or 17%). A high $S_\mathrm{conf}$ for solids suggests partial melting. For example, Figure S 4 shows the predicted $p^n$ values as a function of nearest neighbors $n$ for both Ba and Cl after AIMD simulations at 1300 K using LDA for both solid and liquid $BaCl_2$. The heavy Ba sublattice remains solid with $p^{n(=11)} \approx 1$ and $S_\mathrm{conf}(\mathrm{Ba}) \approx 0$ for the Ba-Ba pairs (Figure S 4e). However, the light Cl sublattice exhibits partial melting with $S_\mathrm{conf}(\mathrm{Cl})$ = 2.70 J/K-mole.atom (Figure S 4g) or 6.15 J/K-mole.atom (Figure S 4h). It is important to note that in the original solid-phase model, only one $n$ value has $p^n = 1$, resulting in $S_\mathrm{conf} = 0$ for each atomic pair, similar to the Ba-Ba case in Figure S 4e but with $p^{n=11} = 1$ and $S_\mathrm{conf} = 0$. Therefore, both the first and the third guidelines are used to estimate $S_\mathrm{conf}$ for Ba and Cl in liquid $BaCl_2$; see Sec. 2.1. For solid $BaCl_2$ and other solid salts, the first guideline is already used to estimate $S_\mathrm{conf}$ by assuming no cross-interaction between the sublattices.

Importantly, excluding configurational entropy, the predicted $T_m$ values are far from experimental data, as demonstrated in the KCl case (Sec. 3.1). For reference, a complete list of these predictions is provided in the supplementary Excel file.

Further analysis reveals that some of the predicted $T_m$ values are sensitive to the choice of exchange-correlation (X-C) functionals such as LDA and GGA. For example, the worst one is for $YF_3$ with its $T_m$ values of 1902 K by LDA and 1190 K by GGA (cf., Figure 5 and Table 2 ), i.e., the predicted $T_m$ is 1546 ± 356 K. Before melting, solid $YF_3$ is partially melted with its $S_{conf}$ values of 2.3 J/K·mol-atom by LDA at an AIMD temperature of 1800 K and 2.0 J/K·mol-atom by GGA at an AIMD temperature of 1500 K (cf., supplementary Excel file). It should be noted that the $T_m$ value is 1428 K for $YF_3$ based on the SSUB database [43] (1660 K from Janz [42]), implying large experimental uncertainty. Other $T_m$ experimental values with notable discrepancies include RbCl, RbF, $SrCl_2$, and $ZnCl_2$ (see Table 2 ), with the suggestions by Janz [42] adopted in the present work.

The same observation as in the KCl case (Sec. 3.1), Figure 5 and Table 2 confirm the general trend: LDA typically overestimates $T_m$ while GGA underestimates it. However, the reverse trend is also observed for several salts such as CsF, $BeCl_2$, $SrF_2$, $BaCl_2$, $BaF_2$, and $Li_2BeF_4$. In addition to $T_m$, the overestimated bonding strength results by LDA results in smaller volumes compared to those from GGA; see supplementary Figure S 5 and Excel file.

A further test is also performed using the metaGGA of r$^2$SCAN [39]. Figure S 6a shows that AIMD simulations using r$^2$SCAN require longer run times — approximately 2 to 3 times longer than those using LDA or GGA. However, Figure S 6b demonstrates that r$^2$SCAN yields improved melting point predictions for $SrF_2$ (better than GGA) and $BaF_2$ (comparable to GGA and experimental values from the SSUB database [43]).

### 3.3 Melting enthalpy and entropy of 25 fluoride and chloride salts

In general, the melting entropy ($\Delta S_m$) and, more notably, the melting enthalpy ($\Delta H_m$) are proportional to the melting point ($T_m$), especially for materials that share the same crystallographic structure. For example, Figure S 3 shows that that both $\Delta H_{\mathrm{m}}$ and $\Delta S_{\mathrm{m}}$ exhibit a strong linear correlation with $T_{\mathrm{m}}$ for the NaCl-type alkali metal fluoride and chloride salts. In contrast, no clear trends are observed for the other salts with different crystallographic structures, as detailed in Table S 1.

Similar to the $T_{\mathrm{m}}$ case, Figure 6 and Figure 7 illustrate the predicted values of melting enthalpy ($\Delta H_{\mathrm{m}}$) and melting entropy ($\Delta S_{\mathrm{m}}$), respectively, for 25 fluoride and chloride salts based on AIMD simulations using both LDA and GGA, compared with experimental data from the SSUB database [43]. All corresponding digital data are provided in Table 2 and the supplementary Excel file.

Figure 6 shows that the predicted $\Delta H_{\mathrm{m}}$ values generally agree well with experimental data from SSUB [43], especially the predictions by LDA. This supports the long-standing view that enthalpy can be reliably predicted by MD simulations [4,7]. The worse $\Delta H_{\mathrm{m}}$ predictions are observed for $YF_3$ by LDA, followed by $SrF_2$ by LDA and $MgCl_2$ by GGA, due possibly to experimental uncertainties (e.g., for $YF_3$, see Sec. 3.2) and the presence of partial melting, as indicated by significant $S_{\mathrm{conf}}$ values in solids. These findings are consistent with the poorer $T_{\mathrm{m}}$ predictions for these salts in Figure 5.

Besides $\Delta H_{\mathrm{m}}$, Figure 7 shows that the predicted $\Delta S_{\mathrm{m}}$ values for these salts agree reasonably well with experimental data from SSUB [43], due largely to the inclusion of configurational entropy; see the KCl case study in Sec. 3.1 and the supplementary Excel file. The worse predictions are observed for CsF by GGA, LiCl by GGA, $YF_3$ by LDA, $YCl_3$ by PBE, $BeF_2$ by LDA, $SrF_2$ by LDA, and $MgCl_2$ by both LDA and GGA. Interestingly, in some cases, accurate $T_{\mathrm{m}}$ predictions observed for these salts, due to compensating errors between $\Delta H_{\mathrm{m}}$ and $\Delta S_{\mathrm{m}}$. For example, the predicted $T_{\mathrm{m}}$ of 968 K for CsF (LDA) closely matches the experimental $T_{\mathrm{m}}$ of 976 K [42,43], despite deviations in $\Delta H_{\mathrm{m}}$ and $\Delta S_{\mathrm{m}}$ components. Conversely, less accurate $\Delta S_{\mathrm{m}}$ values lead to less accurate $T_{\mathrm{m}}$ values, for example, GGA predicts $T_{\mathrm{m}}$ = 644 K for LiCl, compared to the experimental value of 883 K, and LDA predicts $T_{\mathrm{m}}$ = 1092 K for $BeF_2$ versus the experimental 825 K; see Table 2 and the supplementary Excel file.

Despite some limitations in $\Delta H_{\mathrm{m}}$ and $\Delta S_{\mathrm{m}}$ predictions for some salts due to the selected X-C functional, supercell size, and/or the employed nearest-neighbor environments to predict $S_{\mathrm{conf}}$ (cf., **Eq. 6**), the present methodology demonstrates strong capability in accurately predicting entropy, showing good agreement with experimental measurements.

## 4 Summary

The present work demonstrates a novel methodology based on zentropy theory for accurately predicting entropies and entropy-related properties (e.g., melting points) for both solid and liquid phases of 25 binary and ternary fluoride and chloride salts. These salts include cations from alkali metals (Li, Na, K, Rb, and Cs), alkali earth metals (Be, Mg, Ca, Sr, and Ba), and transition metals (Zn and Y). In combination with molecular dynamics (MD) simulations, this approach addresses a long-standing challenge in accurately evaluating configurational entropy in liquids. By leveraging AIMD trajectories, this method yields reliable predictions of entropy, enthalpy, and melting points across a diverse set of molten salts.

Accurate melting points ($T_{\mathrm{m}}$) are predicted for most salts, particularly those with negligible configurational entropy ($S_{\mathrm{conf}}$) in their solids, such as all the NaCl-type alkali metal halides. In contrast, the largest deviations in $T_{\mathrm{m}}$ predictions occur for salts with significant $S_{\mathrm{conf}}$ values in their solids. For example, the predicted $T_{\mathrm{m}}$ error for $BeF_2$ reaches up to 202 K or 24%, followed by notable discrepancies for $BaF_2$ and $SrF_2$. The choice of exchange-correlation (X-C) functionals also significantly affects the predicted $T_{\mathrm{m}}$ values, with LDA generally overestimating and GGA underestimating melting points for most salts. In addition to $T_{\mathrm{m}}$, the predicted melting enthalpy ($\Delta H_{\mathrm{m}}$) and melting entropy ($\Delta S_{\mathrm{m}}$) values also show good agreement with experimental data. The present work highlights the power of zentropy-based simulations for capturing the overlooked configurational entropy in the literature in both liquids and certain solids, using a single MD trajectory. This methodology offers a transformative advancement in computational thermodynamics, enabling accurate, efficient, direct predictions of a broad spectrum of thermodynamic properties, including entropy, enthalpy, Gibbs energy, melting points, and heat capacity, with the potential to extend to properties such as viscosity, interfacial energy, and comprehensive phase diagrams involving liquids.

It is worth noting that the accurately predicting melting points in the present work refers to (1) the use of a rigorous zentropy methodology and (2) the improvement over predictions that omit configurational entropy. To further enhance prediction accuracy, advanced X-C functionals such as the meta-GGA of r$^2$SCAN should be considered, as demonstrated in the present work for $BaF_2$ and $SrF_2$.

## 5 Acknowledgements

This work was funded by U.S. Department of Energy (DOE) with Award No. DE-NE0009288 and the U.S. National Science Foundation (NSF) via Award No. CMMI-2226976. CC and PAB acknowledge, with thanks, support from the Engineering and Physical Sciences Research Council, UK (Grant No. EP/X011607/1). First-principles calculations were performed partially on the Roar supercomputers at the Pennsylvania State University's Institute for Computational and Data Sciences (ICDS), partially on the resources of the National Energy Research Scientific Computing Center (NERSC), a DOE Office of Science User Facility supported under Contract No. DE-AC02-05CH11231 using the NERSC award BES-ERCAP0032760, and partially on the resources of Advanced Cyberinfrastructure Coordination Ecosystem: Services & Support (ACCESS) through allocation DMR1400063, which is supported by U.S. National Science Foundation Grants Nos. 2138259, 2138286, 2138307, 2137603, and 2138296.

## 6 Tables and Table Captions

Table 1. Summary of methods and key equations used herein to predict melting point, enthalpy, entropy, and heat capacity for both solid and liquid phases.

| **Property** | **Method** | **Equation(s)** |
|---|---|---|
| Melting point | Equating Gibbs energies of solid and liquid phases | Eq. 1 |
| Enthalpy | Directly from AIMD, including kinetic and potential energy contributions | |
| Entropy | Vibrational entropy ($S_{vib}$) by phonon analysis using velocity auto-correlation functions from AIMD; electronic entropy ($S_{ele}$, neglected in this work) by AIMD; and configurational entropy ($S_{conf}$) via analysis of the nearest-neighbor environments from AIMD. | Eqs. 3, 6 |
| Heat capacity | Derived from temperature dependence of enthalpy or entropy. | Eqs. 8-9 |

Table 2. Calculated values of melting point ($T_m$), melting enthalpy ($\Delta H_m$), and melting entropy ($\Delta S_m$) of 25 fluoride and chloride salts using the X-C functionals of LDA and GGA, compared with experimental data from Janz [42] and the SGTE substance database (i.e., SSUB) [43]. Note that more details and data from AIMD simulations are provided in the supplementary Excel file.

| No. | Salts | $T_m$ (K) | | | $\Delta H_m$ (kJ/mol-atom) | | | $\Delta S_m$ (J/K·mol-atom) | | |
|---|---|---|---|---|---|---|---|---|---|---|
| | | Expt.[a] | LDA | GGA | SSUB | LDA | GGA | SSUB | LDA | GGA |
| 1 | LiCl | 883 | 1059 | 644 | 9.92 | 10.71 | 5.33 | 11.23 | 10.11 | 8.28 |
| 2 | LiF | 1121 | 1308 | 958 | 13.54 | 14.11 | 9.53 | 12.08 | 10.78 | 9.95 |
| 3 | NaCl | 1074 | 1155 | 910 | 14.08 | 13.62 | 9.70 | 13.11 | 11.80 | 10.66 |
| 4 | NaF | 1269 | 1430 | 1141 | 16.67 | 16.26 | 13.19 | 13.14 | 11.37 | 11.57 |
| 5 | KCl | 1044 | 1130 | 927 | 13.14 | 12.05 | 10.43 | 12.59 | 10.66 | 11.26 |
| 6 | KF | 1131 | 1187 | 1037 | 13.60 | 13.46 | 11.46 | 12.02 | 11.34 | 11.05 |
| 7 | RbCl | **988** [b] | 1047 | 819 | 12.20 | 10.84 | 8.96 | 12.24 | 10.35 | 10.94 |
| 8 | RbF | **1106** [c] | 1123 | 866 | 12.91 | 12.39 | 9.14 | 12.09 | 11.03 | 10.56 |
| 9 | CsCl | 918 | 911 | 854 | 10.19 | 9.27 | 8.69 | 11.09 | 10.17 | 10.16 |
| 10 | CsF | 976 | 968 | 1234 | 10.85 | 10.42 | 7.53 | 11.12 | 10.77 | 6.18 |
| 11 | $BeCl_2$ | 688 | 743 | 850 | 2.89 | 3.84 | 5.56 | 4.20 | 5.17 | 6.55 |
| 12 | $BeF_2$ | 825 | 1092 | 962 | 1.58 | 4.28 | 2.45 | 1.92 | 3.92 | 2.55 |
| 13 | $MgCl_2$ | 987 | 1327 | 918 | 14.37 | 15.09 | 9.82 | 14.56 | 11.38 | 10.77 |
| 14 | $MgF_2$ | 1536 | 1681 | 1523 | 19.57 | 18.19 | 16.68 | 12.74 | 10.82 | 10.95 |
| 15 | $CaCl_2$ | 1048 | 1032 | 1082 | 9.52 | 7.79 | 10.17 | 9.11 | 7.55 | 9.40 |
| 16 | $CaF_2$ | 1691 | 1840 | 1740 | 10.00 | 10.55 | 8.56 | 5.91 | 5.78 | 4.96 |
| 17 | $SrCl_2$ | **1147** [d] | 1175 | 1090 | 7.41 | 4.92 | 5.54 | 6.72 | 4.29 | 5.14 |
| 18 | $SrF_2$ | 1750 | 1358 | 1483 | 9.89 | 4.31 | 7.82 | 5.65 | 3.23 | 5.31 |
| 19 | $BaCl_2$ | 1234 | 1278 | 1544 | 5.28 | 5.12 | 5.80 | 4.28 | 4.13 | 3.84 |
| 20 | $BaF_2$ | 1641 | 1124 | 1605 | 7.79 | 4.80 | 9.08 | 4.74 | 4.32 | 5.69 |
| 21 | $ZnCl_2$ | **563** [e] | 676 | 603 | 3.43 | 4.96 | 3.35 | 5.74 | 7.34 | 5.56 |
| 22 | $ZnF_2$ | 1220 | 1322 | 1000 | 13.33 | 13.68 | 11.11 | 10.93 | 10.35 | 11.11 |
| 23 | $YCl_3$ | 994 | 833 | 858 | 7.87 | 5.50 | 8.90 | 7.91 | 6.61 | 10.45 |
| 24 | $YF_3$ | **1660** [f] | 1902 | 1190 | 6.99 | 14.56 | 8.26 | 4.89 | 7.66 | 6.98 |
| 25 | $Li_2BeF_4$ | 732 | 724 | 791 | 6.28 | 5.90 | 6.13 | 8.58 | 8.16 | 7.75 |

[a] Experimental data by Janz [42] and SSUB [43], showing the same melting points except for the labeled ones (b to f) from Janz, which are used in the present work; [b] 997 K by SSUB; [c] 1068 K by SSUB; [d] 1103 K by SSUB; [e] 598 K by SSUB; [f] 1428 K by SSUB.

## 7 Figures and Figure Captions

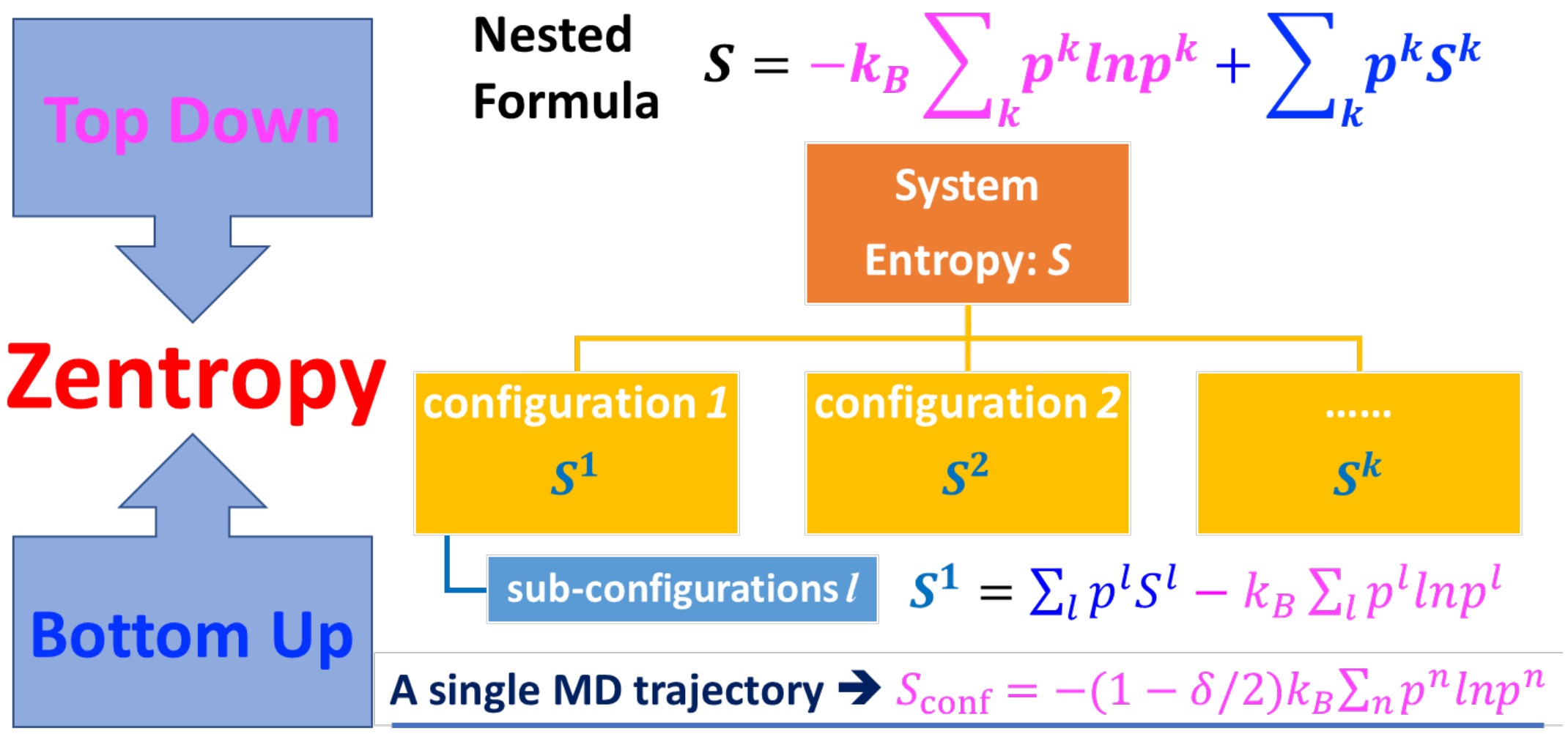

Figure 1. Zentropy represents coarse graining of entropy through nested formulae, spanning from the observed system down to the quantum scale [15–18]. It integrates both a top-down approach, which computes entropy between configurations (highlighted in pink), and a bottom-up approach, which computes entropy within each configuration at the observation scale (highlighted in blue). The present work proposes a novel method to compute the configurational entropy $S_{\text{conf}}$ of solids or liquids using a single molecular dynamics (MD) trajectory by analyzing the overall probabilities $p^n$ for atoms having $n$ nearest neighbors.

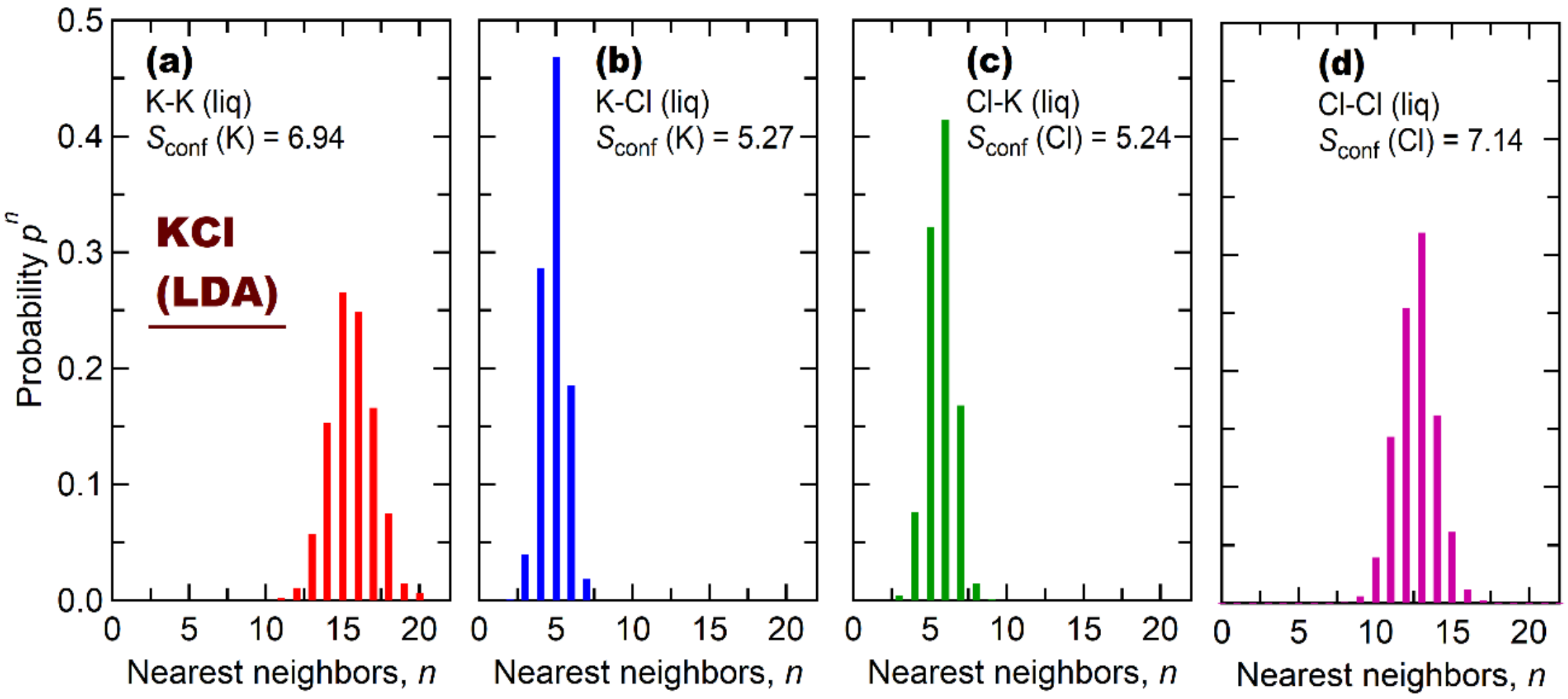

Figure 2. Predicted probability $p^n$ of element K or Cl having $n$ nearest neighbors after AIMD simulations of liquid KCl using LDA: (a) K-K neighbor distribution, (b) K-Cl neighbor distribution, (c) Cl-K neighbor distribution, and (d) Cl-Cl neighbor distribution. The unit of entropy is J/K·mol-atom, and $S_{\mathrm{conf}}(\mathrm{K})$ can be calculated using K-K pairs and/or K-Cl pairs and $S_{\mathrm{conf}}(\mathrm{Cl})$ can be calculated using Cl-K pairs and/or Cl-Cl pairs.

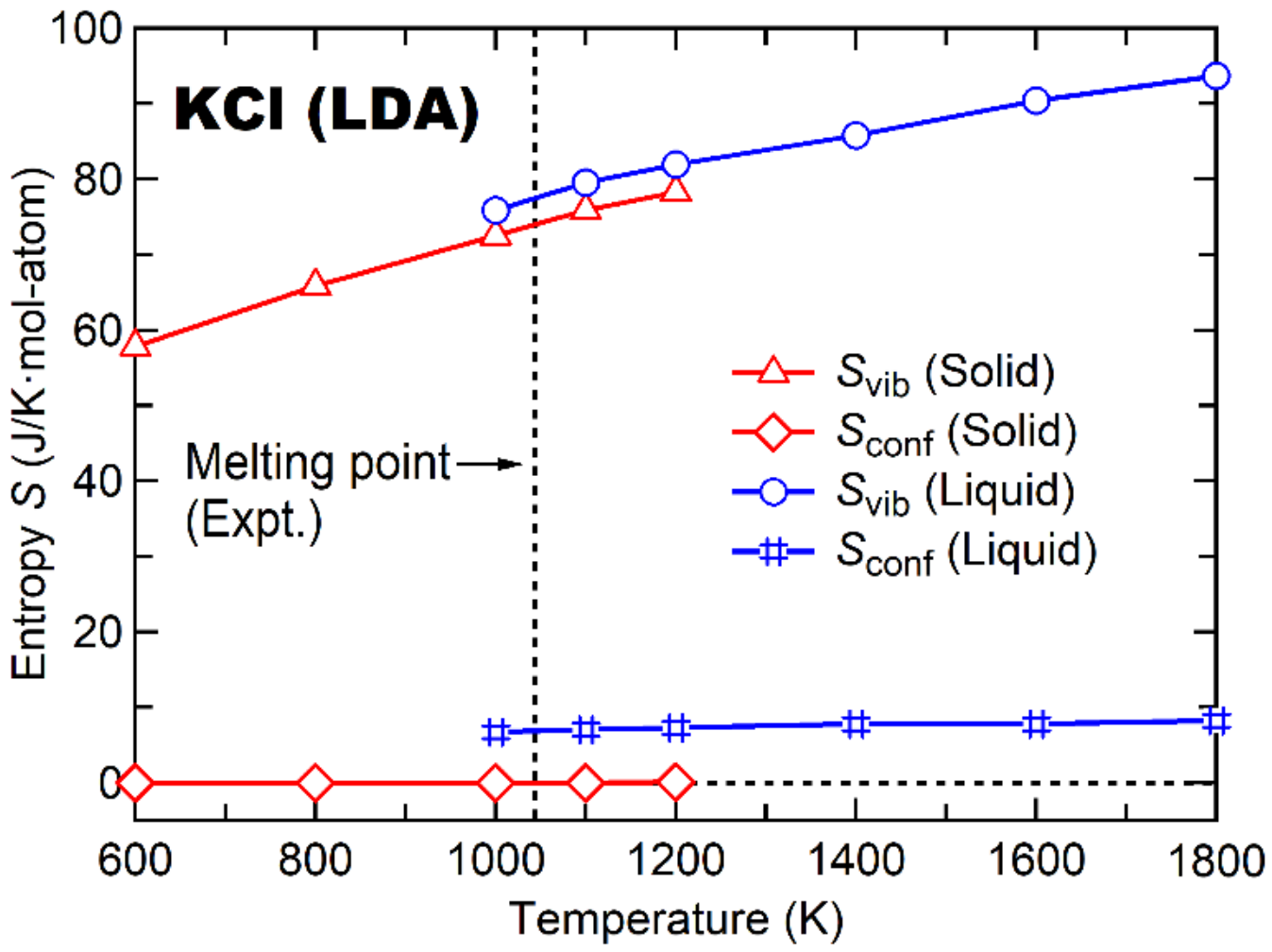

Figure 3. Predicted vibrational entropy $S_{\mathrm{vib}}$ and configurational entropy $S_{\mathrm{conf}}$ as a function of temperature for solid and liquid KCl after AIMD simulations by LDA. Experimental melting point of KCl is 1044 K [42,43].

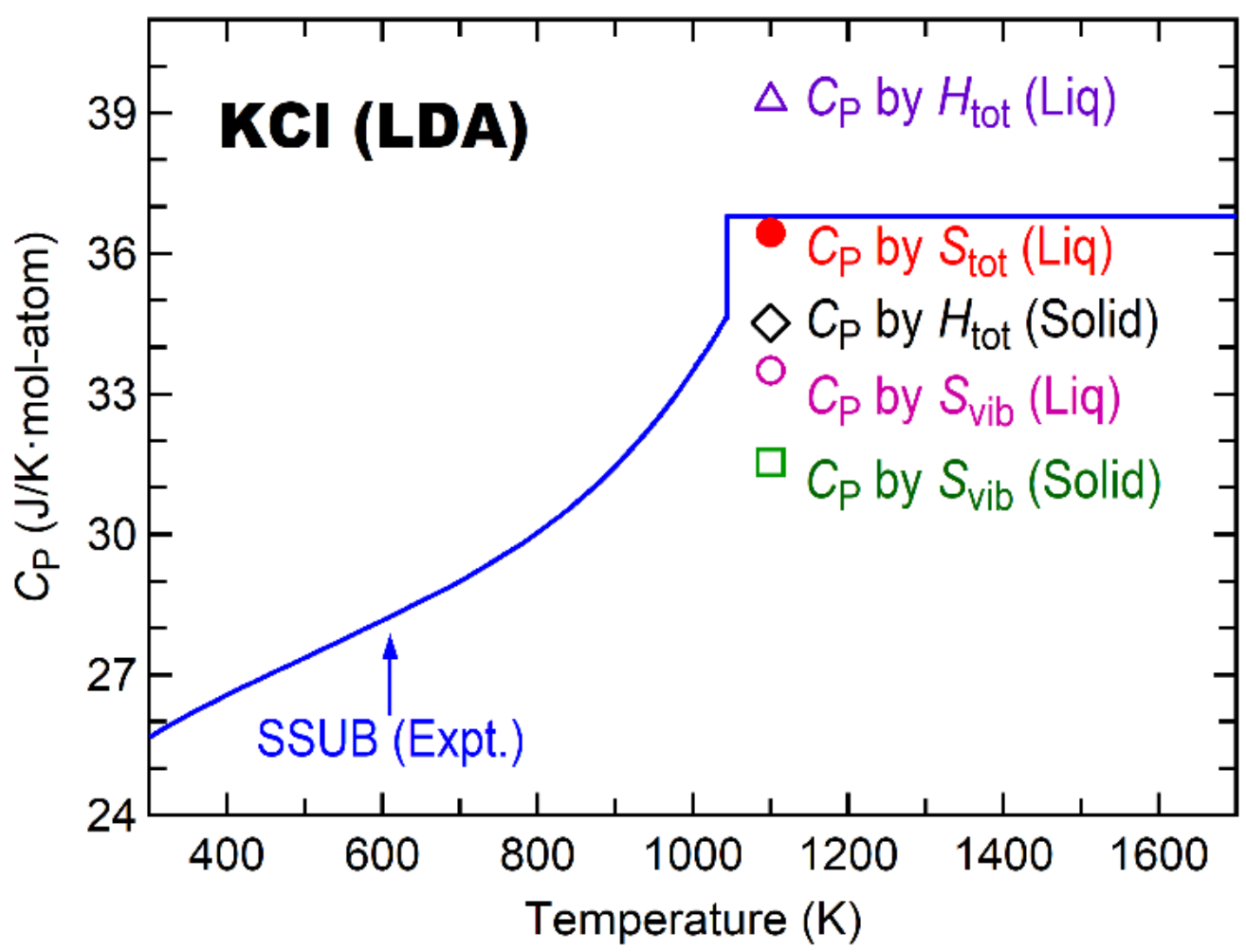

Figure 4. Predicted heat capacity $C_P$ values (the symbols) of liquid and solid KCl based on enthalpy ($H$) and entropy ($S$) after LDA-based AIMD simulations at 1100 K, compared with experimental $C_P$ values based on the SSUB database [43]. The flat $C_P$ above melting point of 1044 K from SSUB arises from the Gibbs energy of liquid KCl being modelled as $a + bT + cTlnT$, where $a$, $b$, and $c$ are modeling parameters.

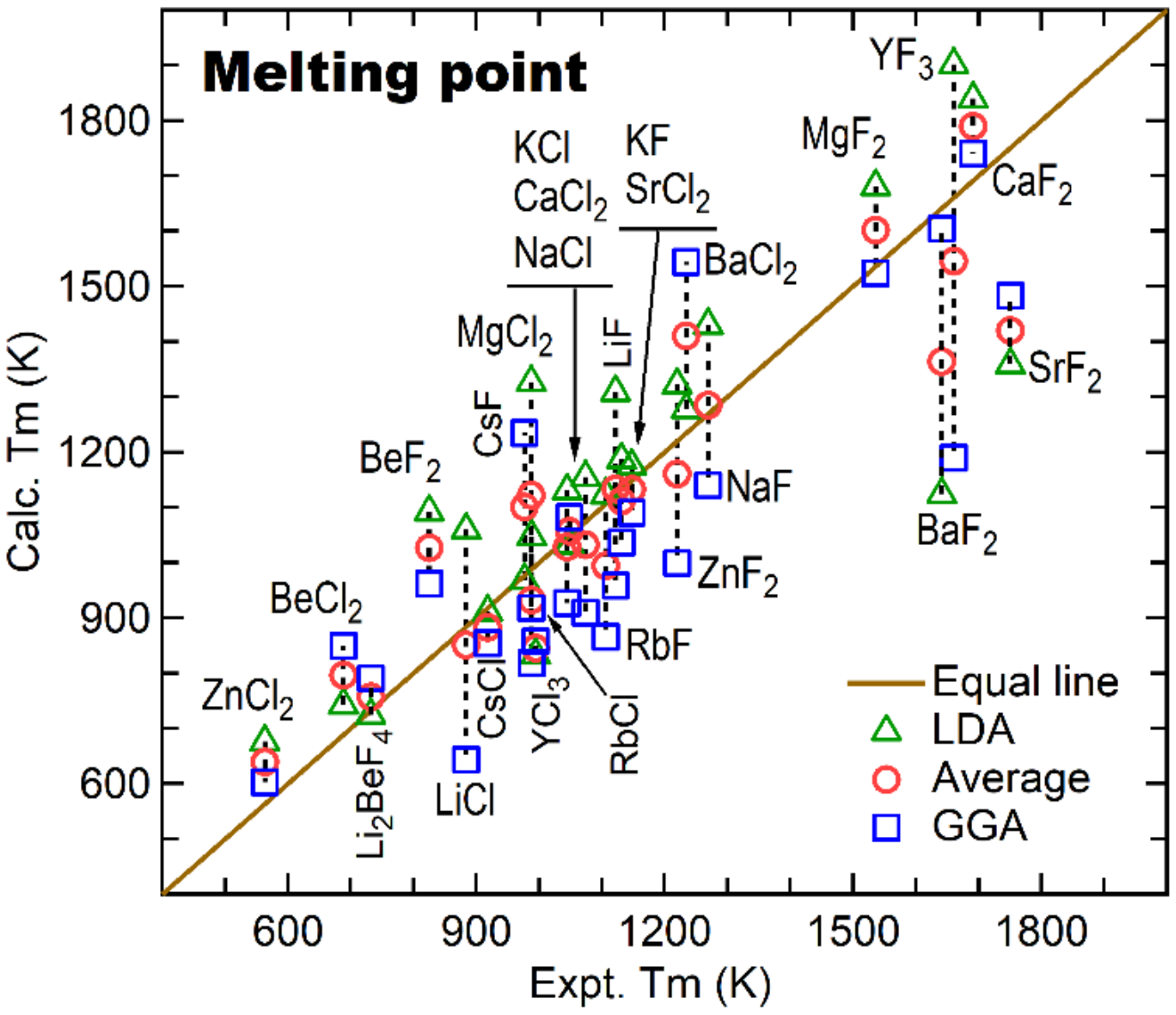

Figure 5. Predicted melting points $T_m$ of 25 Cl- and F-based molten salts using zentropy-based AIMD simulations by LDA and GGA. The digital data are also provided in Table 2 and supplemental Excel file, and the experimental melting points are from Janz [42] and SSUB [43].

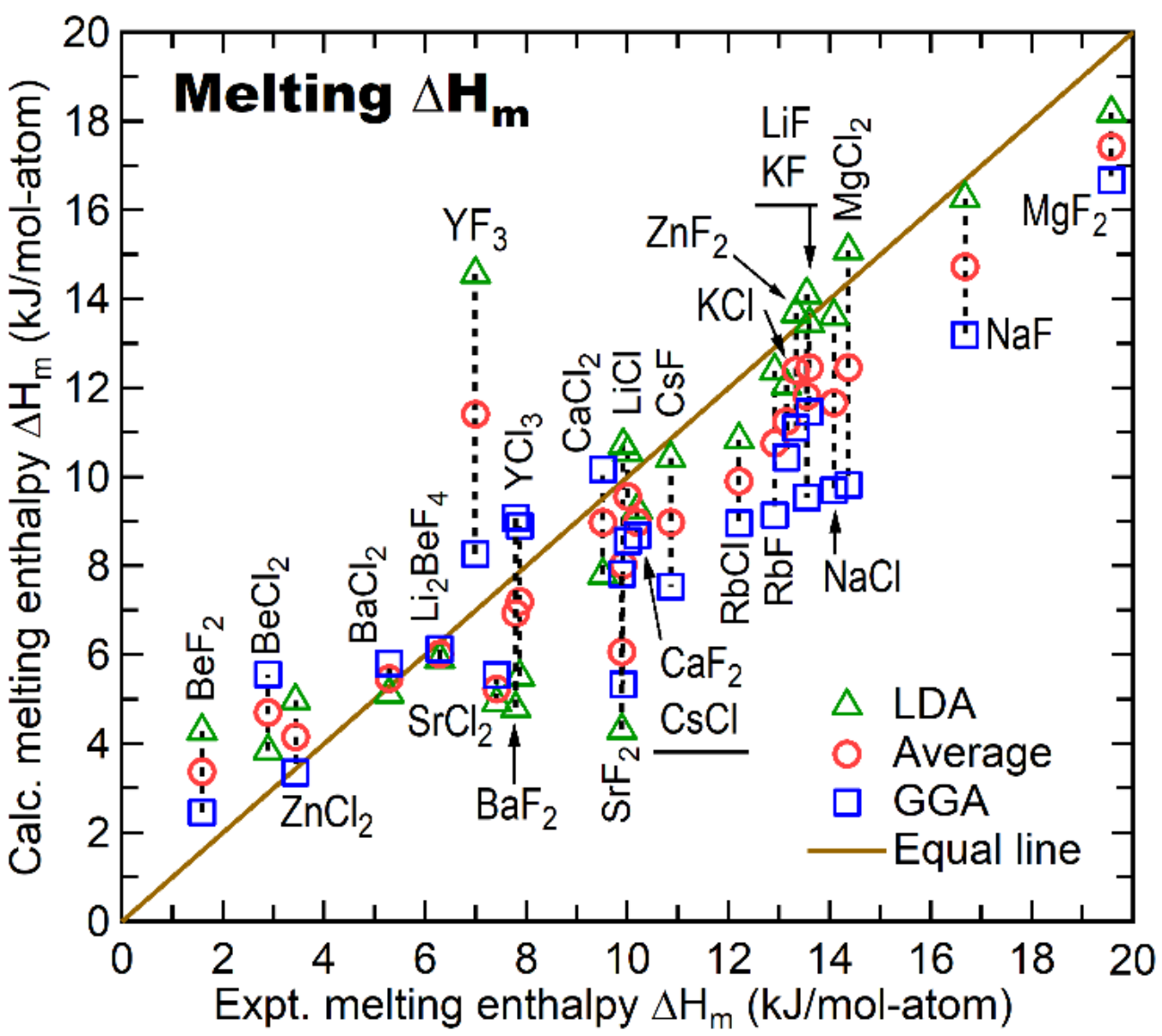

Figure 6. Predicted melting enthalpy $\Delta H_m$ of 25 Cl- and F-based salts using zentropy-based AIMD simulations using LDA and GGA. The digital data are also provided in Table 2 and supplemental Excel file, and the experimental $\Delta H_m$ values are based on SSUB [43].

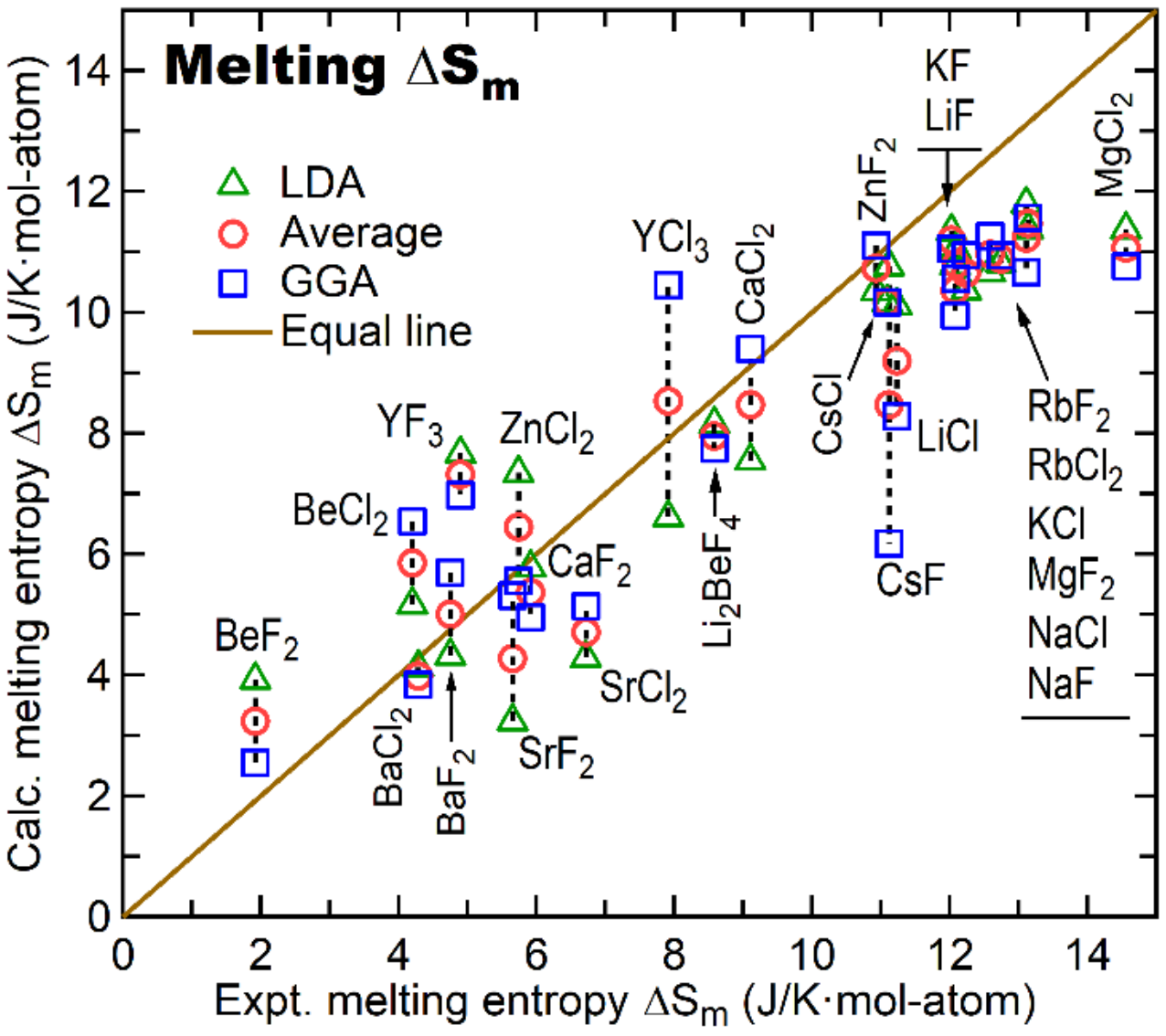

Figure 7. Predicted melting entropy $\Delta S_m$ of 25 Cl- and F-based salts using zentropy-based AIMD simulations using LDA and GGA. The digital data are also provided in Table 2 and supplemental Excel file, and the experimental $\Delta S_m$ values are based on SSUB [43].

# Supplementary Material:

# Achieving accurate entropy and melting point by *ab initio* molecular dynamics and zentropy theory: Application to fluoride and chloride molten salts

Shun-Li Shang,[1,*] Nigel L. E. Hew,[1] Rushi Gong,[1] Cillian Cockrell,[2] Paul A. Bingham,[3] Xiaofeng Guo,[4] Jingjing Li,[5] Qi-Jun Hong,[6] and Zi-Kui Liu[1]

1. Department of Materials Science and Engineering, The Pennsylvania State University, University Park, PA 16802, USA
2. Nuclear Futures Institute, Bangor University, Bangor LL57 1UT, UK
3. School of Engineering and Built Environment, Sheffield Hallam University, Sheffield S1 1WB, UK
4. Department of Chemistry, Washington State University, Pullman, Washington, 99164, USA
5. Department of Industrial and Manufacturing Engineering, The Pennsylvania State University, University Park, PA 16802, USA
6. Materials Science and Engineering, School for Engineering of Matter, Transport, and Energy, Arizona State University, Tempe, Arizona 85285, USA

*E-mail: sus26@psu.edu (S. L. Shang)

## One Supplementary Table

Table S 1. Crystallographic structures of 25 fluoride and chloride salts used in the present work, see also their structure details in Materials Project (mp) [1] and Open Quantum Materials Database (OQMD) [2].

| No. | Salts | Space group | Structure details |
|---|---|---|---|
| 1 | LiCl | $Fm\bar{3}m$ | NaCl-type, B1 structure, see also mp-22905 for LiCl |
| 2 | LiF | $Fm\bar{3}m$ | NaCl-type, B1 structure, see also mp-1138 for LiF |
| 3 | NaCl | $Fm\bar{3}m$ | NaCl-type, B1 structure, see also mp-22862 for NaCl |
| 4 | NaF | $Fm\bar{3}m$ | NaCl-type, B1 structure, see also mp-682 for NaF |
| 5 | KCl | $Fm\bar{3}m$ | NaCl-type, B1 structure, see also mp-23193 for KCl |
| 6 | KF | $Fm\bar{3}m$ | NaCl-type, B1 structure, see also mp-463 for KF |
| 7 | RbCl | $Fm\bar{3}m$ | NaCl-type, B1 structure, see also mp-23295 for RbCl |
| 8 | RbF | $Fm\bar{3}m$ | NaCl-type, B1 structure, see also mp-11718 for RbF |
| 9 | CsCl | $Fm\bar{3}m$ | NaCl-type, B1 structure, see also mp-573697 for CsCl |
| 10 | CsF | $Fm\bar{3}m$ | NaCl-type, B1 structure, see also mp-1784 for CsF |
| 11 | $BeCl_2$ | $Ibam$ | $SiS_2$-type, see also mp-23267 or OQMD-6092 for $BeCl_2$ |
| 12 | $BeF_2$ | $P3_121$ | α-SiO2-type, see also mp-15951 or OQMD-7499 for $BeF_2$ |
| 13 | $MgCl_2$ | $R\bar{3}m$ | $CdCl_2$-type, see also mp-23210 or OQMD-53639 for $MgCl_2$ |
| 14 | $MgF_2$ | $P4_2/mnm$ | Rutile-type, see also mp-1249 or OQMD-1697 for $MgF_2$ |
| 15 | $CaCl_2$ | $P4_2/mnm$ | Rutile-type, see also mp-22904 or OQMD-10159 for $CaCl_2$ |
| 16 | $CaF_2$ | $Fm\bar{3}m$ | $CaF_2$-type, see also mp-2741 or OQMD-5585 for $CaF_2$ |
| 17 | $SrCl_2$ | $Fm\bar{3}m$ | $CaF_2$-type, see also mp-23209 or OQMD-3681 for $SrCl_2$ |
| 18 | $SrF_2$ | $Fm\bar{3}m$ | $CaF_2$-type, see also mp-981 or OQMD-7257 for $SrF_2$ |
| 19 | $BaCl_2$ | $Fm\bar{3}m$ | $CaF_2$-type, see also mp-568662 or OQMD-2092 for $BaCl_2$ |
| 20 | $BaF_2$ | $Fm\bar{3}m$ | $CaF_2$-type, see also mp-1029 or OQMD-647336 for $BaF_2$ |
| 21 | $ZnCl_2$ | $I\bar{4}2d$ | See mp-22909 or OQMD-3263 for $ZnCl_2$ |
| 22 | $ZnF_2$ | $P4_2/mnm$ | Rutile-type, see also mp-1873 or OQMD-2482 for $ZnF_2$ |
| 23 | $YCl_3$ | $C2/m$ | $RhBr_3$-type, see also mp-27455 or OQMD-3179 for $YCl_3$ |
| 24 | $YF_3$ | $Pnma$ | Cementite structure, see also mp-2416 or OQMD-2023031 for $YF_3$ |
| 25 | $Li_2BeF_4$ | $R\bar{3}$ | $Be_2SiO_4$-type structure, see also mp-4622 or OQMD-2927 for $Li_2BeF_4$ |

## Six Supplementary Figures

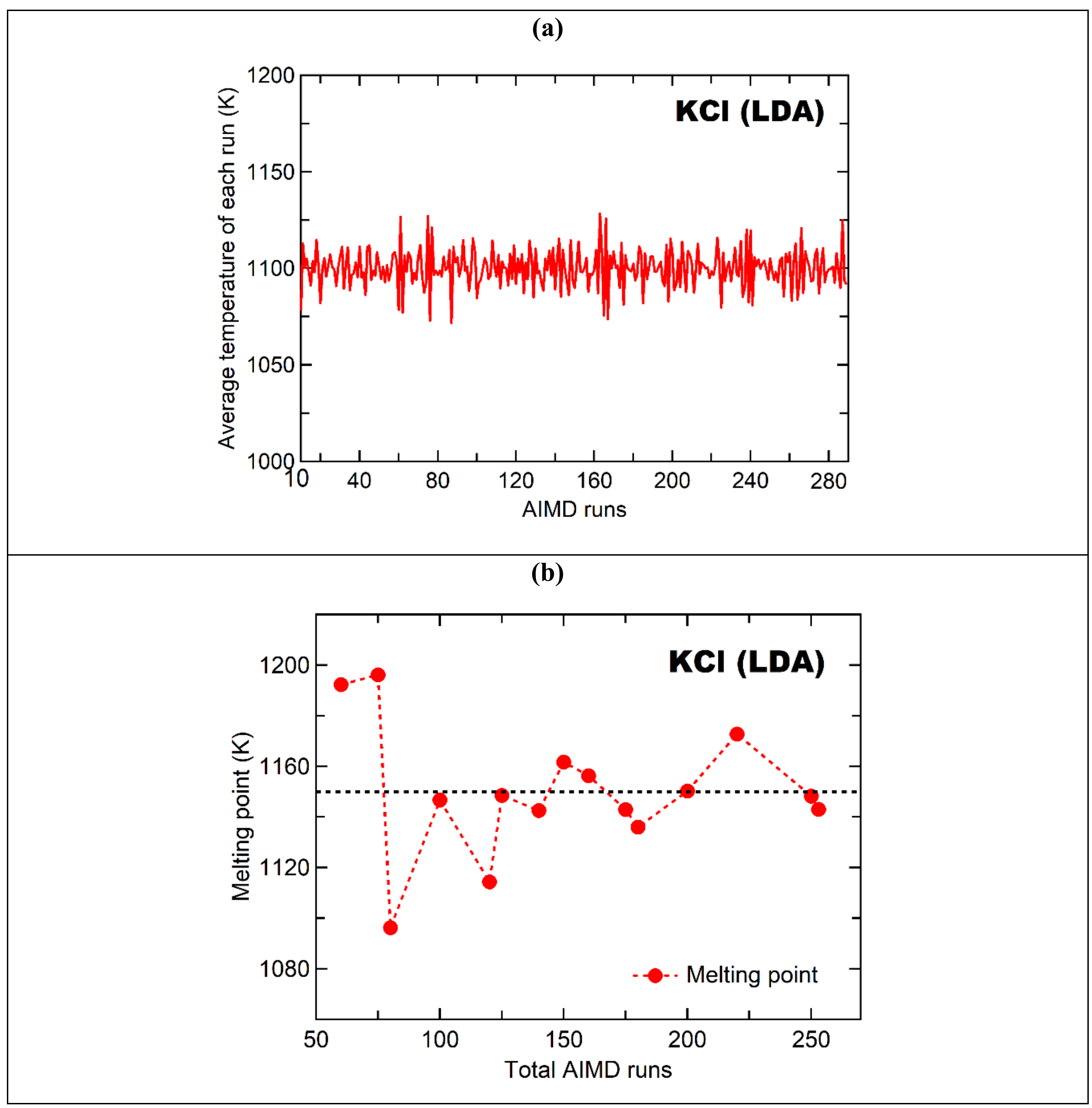

Figure S 1. Test of KCl by LDA at 1100 K. (a) Predicted average temperature of each AIMD run of liquid KCl. At each run, the standard deviation is 90 K, and after 10 AIMD runs, the simulations reach the targeted temperature of 1100 K, i.e., the prediction is 1100 ± 10 K. (b) Predicted melting point ($T_m$) of KCl by LDA as a function of the number of total AIMD runs at 1100 K. The first 50 runs were discarded to ensure equilibration and are excluded from the $T_m$ predictions. After approximately 100 runs, the predicted $T_m$ stabilizes around 1150 K, with an average error of ±10 K (or 1%). Note that each AIMD run consists of 80 ionic steps.

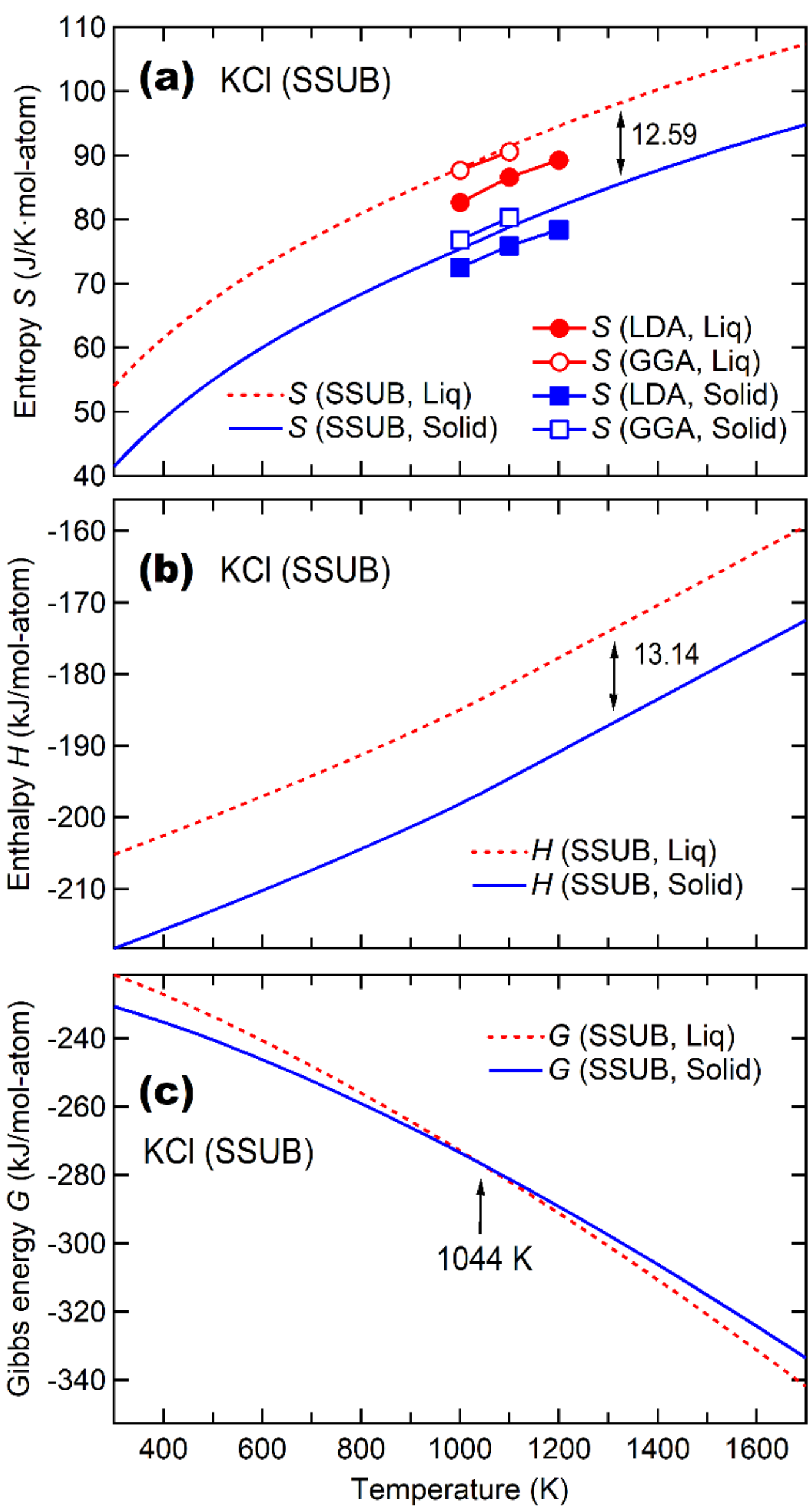

Figure S 2. Calculated values of entropy (a), enthalpy (b), and Gibbs energy (c) of liquid and solid KCl based on SSUB [3]. The present predictions of total entropy for liquid and solid KCl using LDA and GGA are also shown for comparison.

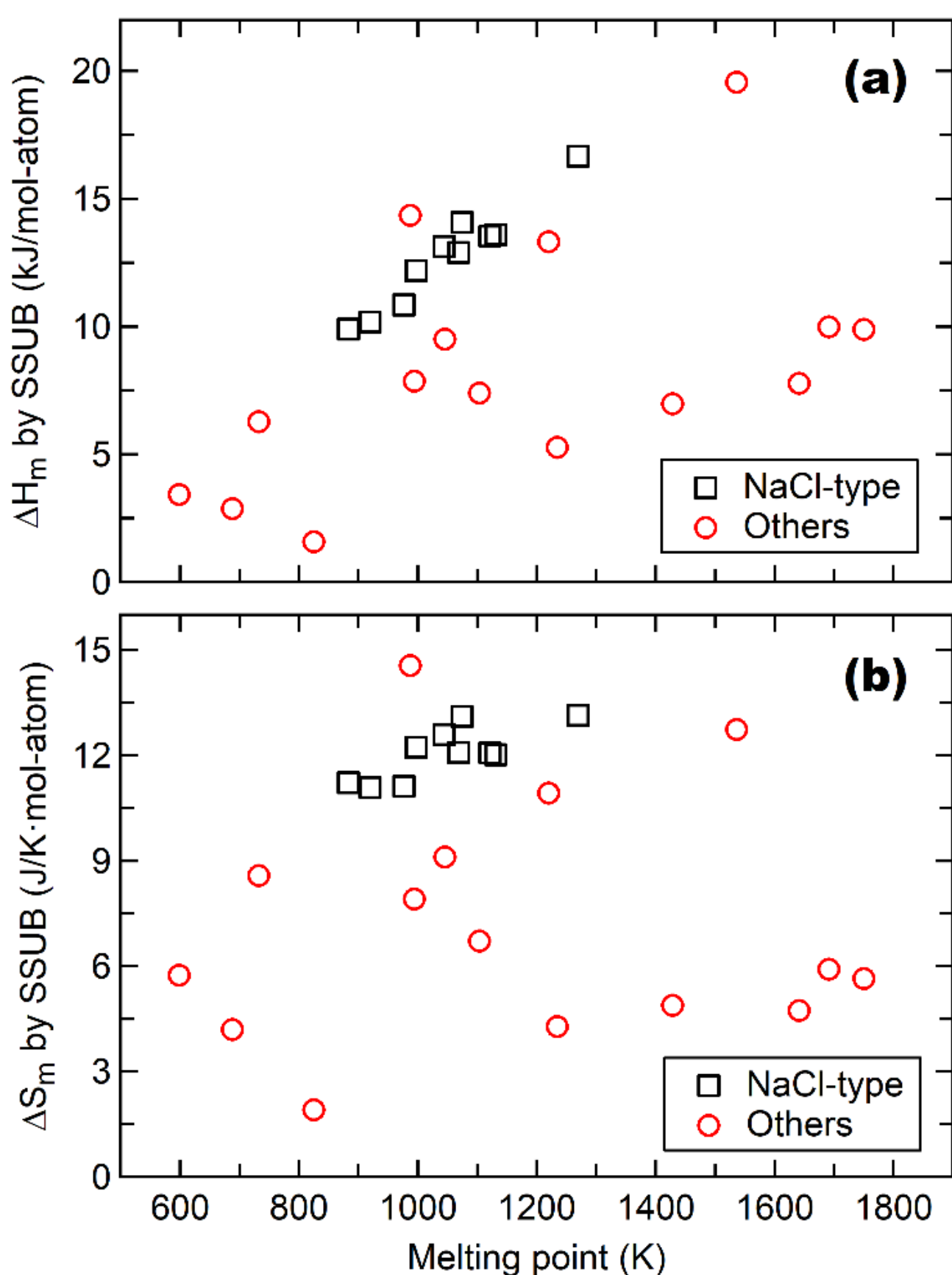

Figure S 3. Melting enthalpy $\Delta H_\mathrm{m}$ (a) and melting entropy $\Delta S_\mathrm{m}$ (b) of 25 Cl- and F-based salts as a function of melting point based on SSUB [3], where the digital data are provided in supplemental Excel file.

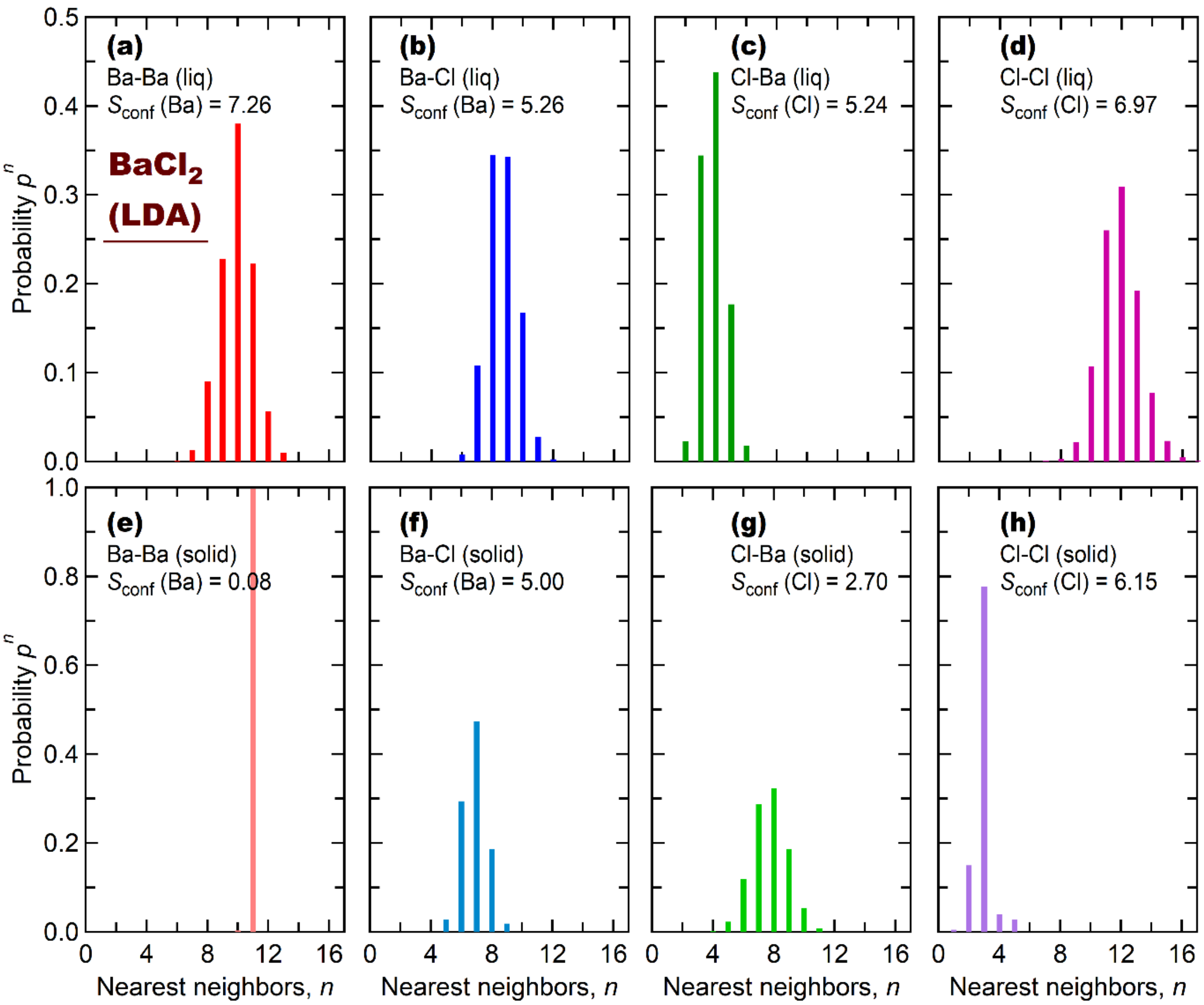

Figure S 4. Predicted probability $p^n$ of element Ba or Cl having $n$ nearest neighbors after AIMD simulations of liquid and solid $BaCl_2$ using LDA: (a) – (d) for liquid $BaCl_2$ and (e) – (h) for solid $BaCl_2$. The unit of entropy is J/K·mol-atom, and $S_{\mathrm{conf}}(\mathrm{Ba})$ can be calculated using Ba-Ba pairs and/or Ba-Cl pairs and $S_{\mathrm{conf}}(\mathrm{Cl})$ can be calculated using Cl-Ba pairs and/or Cl-Cl pairs.

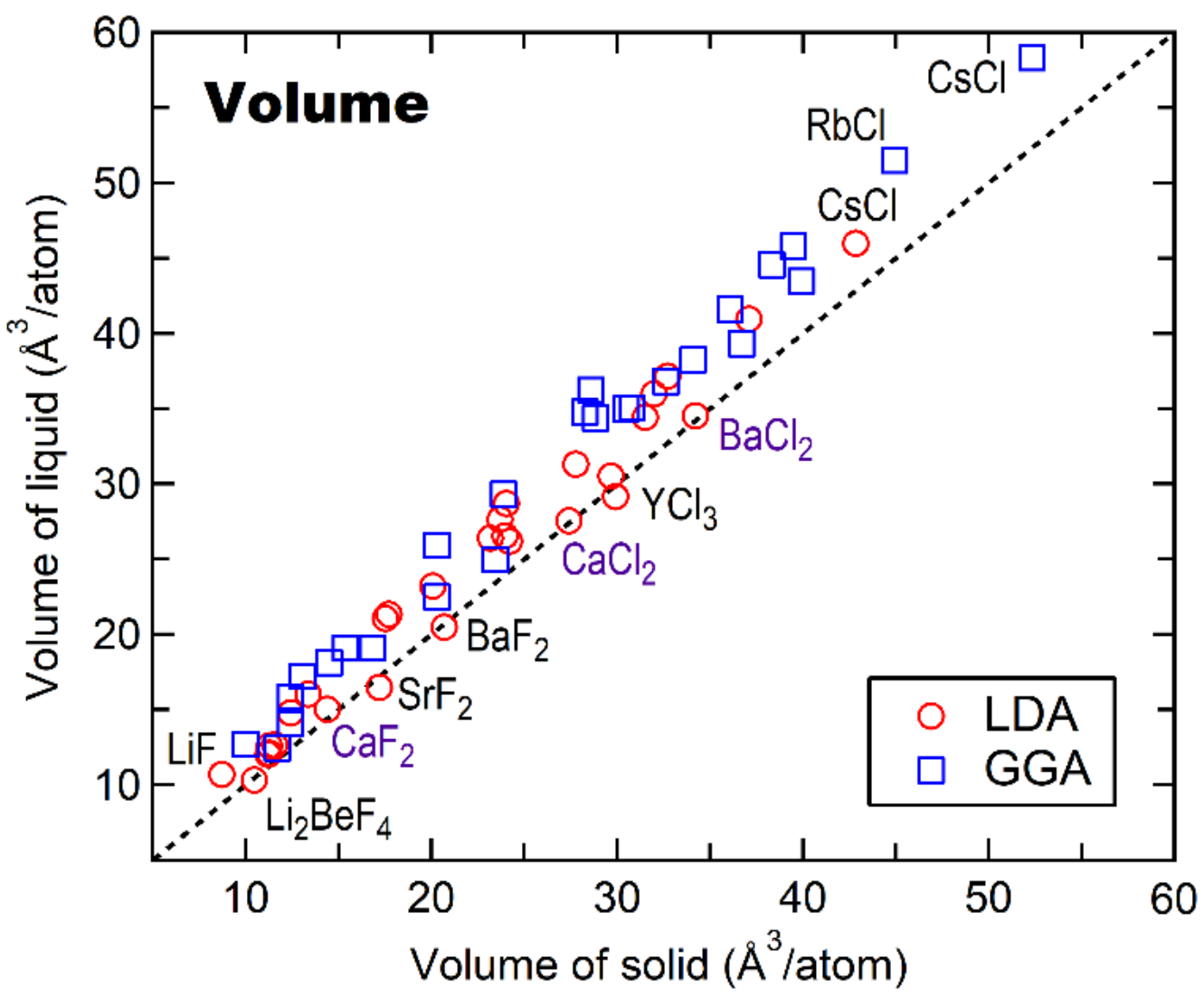

Figure S 5. Predicted equilibrium volumes for 25 solid and liquid salts by AIMD simulations using LDA and GGA, see the digital data given in the supplementary Excel file. Note that LDA predicts that liquid volumes are identical and even less than solid volumes for seven salts (marked, below the dashed line).

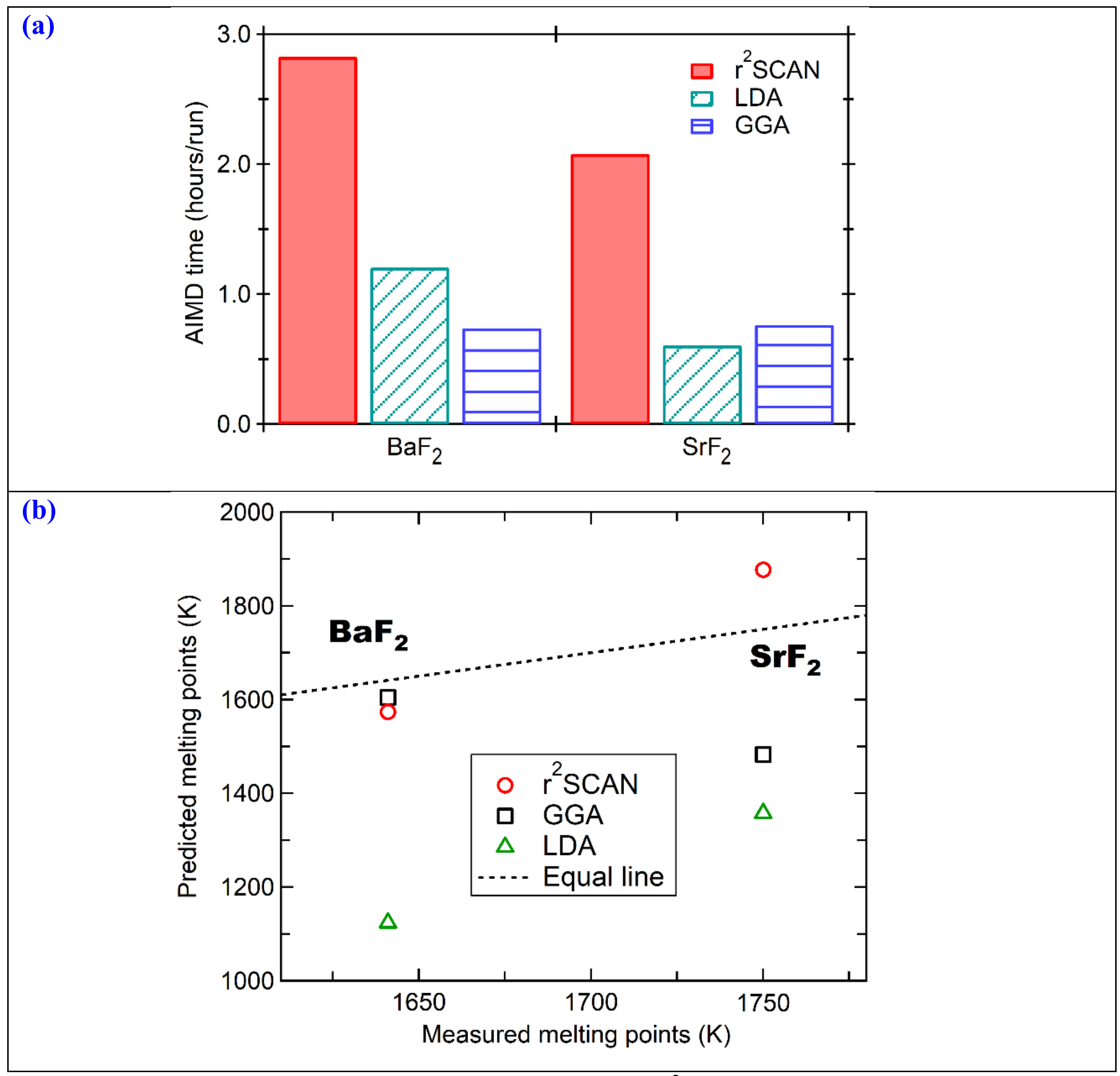

Figure S 6. Test calculations using the meta-GGA of r$^2$SCAN [4] for $BaF_2$ at 1500 K and for $SrF_2$ at 1600 K; see also detailed results in the supplementary Excel file. (a) Average running time for each AIMD run (80 ionic steps) using 32 CPUs at the Bridges-2 supercomputer (https://www.psc.edu/resources/bridges-2), and (b) predicted melting points by GGA, LDA, and r$^2$SCAN in comparison with experimental values based on the SSUB database [3].